\documentstyle[epsfig]{elsart}

\def\spose#1{\hbox to 0pt{#1\hss}}
\def\ltapprox{\mathrel{\spose{\lower 3pt\hbox{$\mathchar"218$}}
 \raise 2.0pt\hbox{$\mathchar"13C$}}}
\def\gtapprox{\mathrel{\spose{\lower 3pt\hbox{$\mathchar"218$}}
 \raise 2.0pt\hbox{$\mathchar"13E$}}}
\def\inapprox{\mathrel{\spose{\lower 3pt\hbox{$\mathchar"218$}}
 \raise 2.0pt\hbox{$\mathchar"232$}}}

\def\dfrac#1#2{{\displaystyle {#1 \over #2}}}
\def\slash#1{\mbox{$\not \!\! #1$}}
\newcommand{\de}{\partial}

\newcommand{\Pj}{\mbox{I}\!\!\mbox{P}}

\newcommand{\<}{\langle}
\renewcommand{\>}{\rangle}
\newcommand{\Tr}{\mbox{Tr}\;} 
\renewcommand{\crit}{\mbox{\scriptsize crit}}
\newcommand{\latt}{\mbox{\scriptsize latt}}
\newcommand{\BPT}{\mbox{\scriptsize BPT}}
\newcommand{\QCD}{\mbox{\scriptsize QCD}}
\newcommand{\RGI}{\mbox{\scriptsize RGI}}
\newcommand{\RI}{\mbox{\scriptsize RI}}

\newcommand{\msbar}{\overline{\mbox{\scriptsize MS}}}
\newcommand{\MSbar}{\overline{\mbox{MS}}}

\newcommand{\bea}{\begin{eqnarray}}
\newcommand{\eea}{\end{eqnarray}}

\newcommand{\dslash}{\slash{D}}

\newcommand{\lp}{\left(}
\newcommand{\rp}{\right)}

\begin{document}

\rightline{Edinburgh 97/24}
\rightline{FTUV/98-44}
\rightline{IFIC/98-45}
\rightline{ROME1-1184/97}
\rightline{SNS/PH/1998-010}

\begin{frontmatter}

\title{Non-perturbative Renormalization of Quark bilinears}

\author{V.~Gim\'enez}
\address{Dep. de Fisica Teorica and IFIC, Univ. de Val\`encia,\\
Dr. Moliner 50, E-46100, Burjassot, Val\`encia, Spain}
\author{L.~Giusti}
\address{Scuola Normale Superiore, P.zza dei Cavalieri 7, I-56100 Pisa, 
Italy\\  INFN Sezione di Pisa, I-56100 Pisa, Italy}
\author{F.~Rapuano}
\address{Dipartimento di Fisica, Universit\`a di Roma \lq La Sapienza\rq ~and\\
         INFN, Sezione di Roma, P.le A. Moro 2, I-00185 Roma, Italy.}
\author{M.~Talevi}
\address{Department of Physics \&\ Astronomy, University of Edinburgh\\
         The King's Buildings, Edinburgh EH9 3JZ, UK}

\begin{abstract} 
We compute non-perturbatively the renormalization constants of quark bilinears 
on the lattice in the quenched approximation at three values of the coupling 
$\beta=6/g_0^2=6.0,6.2,6.4$ 
using both the Wilson and the tree-level improved SW-Clover fermion action. 
We perform a Renormalization Group analysis at the 
next-to-next-to-leading order and compute Renormalization Group invariant 
values for the constants.
The results are applied to obtain a fully non-perturbative estimate of the
vector and pseudoscalar decay constants.  
\end{abstract}

\end{frontmatter}

\vfill
\centerline{PACS: 11.15.H, 12.38.Gc, 13.30.Eg, 14.20.-c and 14.40.-n}

\newpage
\clearpage 
\pagestyle{plain} \setcounter{page}{1}

\section{Introduction}
Lattice QCD is an unique tool to compute non-perturbatively from first
principles the mass spectrum, leptonic decay constants and in general 
hadronic matrix elements of local operators. 
Renormalization constants, relating the operators
on the lattice to the continuum are necessary 
to extract physical informations from Monte Carlo simulations.
In this paper we study the renormalization properties of composite bilinear
operators \cite{Bochicchio,MM} with the quark action discretized 
{\em a l\'a} Wilson.  

In principle, the renormalization of quark bilinears can be computed
in one-loop perturbation theory, as there are no power divergences 
\cite{Zhang,Borrelli}. 
It is well known, though, that lattice perturbation theory is 
ill behaved, due to the presence of tadpole-like diagrams \cite{Parisi,Lepage}
and at values of the coupling $\beta=6/g_0^2=6.0-6.4$, the 
higher-order corrections may not be small, thus introducing a
large uncertainty in the calculation of the renormalized matrix elements 
in some continuum scheme.  
These problems are avoided using non-perturbative (NP) 
renormalization techniques \cite{NPM,Jansen}. The procedure proposed in 
\cite{NPM} allows a full non-perturbative computation of the matrix elements 
of composite operators in the Regularization Independent (RI) scheme 
\cite{NPM,eps'/eps}. The matching between the RI scheme and $\MSbar$, which is
intrinsically perturbative, is computed using only continuum perturbation 
theory, which is well behaved. 
This method has been shown to be quite successful in reproducing results 
obtained by other methods, such as chiral Ward Identities \cite{Ward}.
Quite impressive is also the influence of the NP renormalization
in the measurement of the quark masses \cite{m_q_APE}, the
chiral condensate \cite{chi} and,
for four-fermion operators, in the restoration of the correct chiral behaviour
of the $B$-parameters in weak decays \cite{DS=2,BK78,BK-JLQCD}.  
Moreover, any attempt to tackle the question
of the $\Delta I=1/2$ rule must rely on NP methods \cite{DI=1/2}.

In this paper, we extend the exploratory computations 
done in \cite{NPM} to a high statistics study with both the standard 
Wilson action and the tree-level improved SW-Clover action \cite{SW}
at different values of the coupling in the quenched approximation.
We compare the dependence on the renormalization scale of the renormalized 
operators with respect to the solution predicted from the Renormalization 
Group Equation (RGE) at the next-to-next-to-leading-order (NNLO).   
Moreover, we use this analysis to estimate the systematics due to
discretization errors on the renormalization constants.

Recently, there has been much progress in the Symanzik on-shell improvement
program \cite{Symanzik}, obtaining a non-perturbative determination of
the renormalization constants with a fully $O(a)$ improved action 
\cite{Alpha,m_q_SF}.   An extension of this program, to take into account
all terms of $O(a)$ including the ones proportional to the quark mass,
has been proposed in \cite{NPI}.  Whereas the non-perturbative improvement
is the direction of future, 
it is true that Wilson and tree-level Clover actions are 
still widely used, e.g. for four-fermion operators relevant to weak decays
(cf. for example ref.~\cite{Sharpe} and references therein).    
Moreover, the study of the renormalization properties in the chiral limit
with an unimproved and partially improved action will provide further
insight for the discretization effects in lattice QCD.

The outline of this paper is as follows.  In sec.~\ref{sec:NPM} we review
the non-perturbative method (NPM) proposed in ref.~\cite{NPM} 
and set the notation for the remainder of this work.  
The Renormalization Group (RG) analysis of the quark bilinears is
outlined in sec.~\ref{sec:RGE}, while in sec.~\ref{sec:Z} we present 
the numerical results for the renormalization constants and discuss the 
systematic errors.  In sec.~\ref{sec:matrix} we apply these results to the
the computation of the leptonic decay constants of the vector and
pseudoscalar mesons.  We finish with our
conclusions in sec.~\ref{sec:conclusions}.

\section{Non-perturbative renormalization method}
\label{sec:NPM}

In this section we review the method of ref.~\cite{NPM}, which we have 
used to compute non-perturbatively the renormalization constants of quark 
bilinears in the Regulatization Independent (RI) scheme \cite{eps'/eps}.  
The method imposes renormalization conditions 
non-perturbatively, directly on quark and gluon Green functions, in a fixed
gauge, with given off-shell external states of large virtuality. Notice that
in RI the renormalization conditions are independent of the regularization 
scheme but they depend on the external states and on the gauge used in the 
procedure.

The renormalization scale $\mu^2$, determined from the virtuality of the 
external states $p^2$, must satisfy the condition 
$\Lambda_{\QCD}\ll\mu\ll O(1/a)$, see \cite{NPM}.  

We have worked in the lattice Landau gauge, defined by minimizing
the functional 
\begin{equation}
\Tr~\left[\sum_{\mu=1}^4(U_{\mu}(x)+U^{\dag}_{\mu}(x))\right]\; .
\end{equation}
The necessity to fix the gauge introduces a systematic uncertainty
due to the existence of both continuum and lattice Gribov copies 
\cite{Parrinello} and the numerical noise that they can generate.
These effects are expected to die off at large virtuality and  
on the renormalization of two-quark operators have been found to be 
small, comparable to the statistical noise \cite{Paciello}.
We are making the assumption that the
Landau lattice gauge-fixing procedure brings gauge fixed lattice operators
into the corresponding continuum ones as $a\to 0$ \cite{Giusti}. 

Let us consider a local lattice quark bilinear 
$O_{\Gamma}=\bar q\Gamma q$, where $\Gamma$ is a Dirac matrix\footnote{In
the following, we shall denote with $\Gamma=A,V,P,S$ the axial and vector 
currents and the pseudoscalar and scalar densities.}.
The renormalization condition is imposed on the 
amputated Green function computed between off-shell quark states of 
momentum $p$ in the Landau gauge 
\begin{equation}
\Lambda_\Gamma(pa)=S_q(pa)^{-1}G_\Gamma(pa)S_q(pa)^{-1}
\end{equation}  
where $G_\Gamma(pa)$ and $S_q(pa)$ are the non-amputated Green function and
quark propagator, calculated non-perturbatively via Monte Carlo simulations
\cite{NPM}.
The renormalization constant $Z_\Gamma^{\RI}(\mu a, g_0)$ of $O_\Gamma$, in the
RI scheme, is determined by the condition
\begin{equation}
Z_\Gamma^{\RI}(\mu a)Z_q^{-1}(\mu a) \Tr\Pj_\Gamma\Lambda_\Gamma (pa)|_{p^2=\mu^2}=1,
\label{eq:RI}
\end{equation}
and the renormalized operator is related to the bare one by
$\hat O_\Gamma^{\RI} =Z_\Gamma^{\RI} O_\Gamma $.  In eq.~(\ref{eq:RI}) 
$\Pj_\Gamma$ is a suitable projector on the tree-level amputated
Green function.  In the case of the quark bilinears the 
projector is simply proportional to $\Gamma^{\dag}$.
$Z_q$ is the wave function renormalization which can be defined
from the Ward Identity (WI) as \cite{NPM}
\begin{equation}
Z_q(\mu a)=-i
\frac{1}{12} Tr \left[\frac{\de S(pa)^{-1}}{\de\slash{p}}\right]\left.
\right|_{p^2=\mu^2}\; .
\label{eq:Z_q_WI} 
\end{equation}
To avoid derivatives with respect to a discrete variable, we have used 
\begin{equation}
Z'_q(\mu a)=\left.-i\frac{1}{12}\frac{Tr \sum_{\mu=1,4}
\gamma_\mu \sin(p_\mu a)S(pa)^{-1}}
{4\sum_{\mu=1,4}\sin^2(p_\mu a)}\right|_{p^2=\mu^2}\; ,
\label{eq:Z_q'_WI} 
\end{equation}
which, in the Landau Gauge, differs from $Z_q$ by a finite term of order
$\alpha_s^2$. The matching coefficient can be computed using continuum
perturbation theory only, and up to order $\alpha_s^2$ \cite{mass_NNLO} 
\begin{equation}
\frac{Z_q}{Z_q^\prime}=
1-\dfrac{\alpha_s^2}{\left( 4\pi \right) ^2}\Delta _q^{(2)}+\ldots
\label{eq:deltaq1}
\end{equation}
with, in the Landau gauge,
\begin{equation}
\Delta^{(2)}_q =  
 {{\left( N_c^2 - 1 \right) }\over {16\,N_c^2}} \,
 \left(3 + 22 N_c^2 - 4 N_c n_f \right)\, .
\label{eq:deltaq2}
\end{equation}
where $N_c$ is the number of colours and $n_f$ the number of active quarks. 
Eqs.~(\ref{eq:RI}) and (\ref{eq:Z_q'_WI}) define the constants 
$Z^{'\RI}_\Gamma$.
From eq.~(\ref{eq:deltaq1}) we obtain for $Z^{\RI}_\Gamma$ 
\begin{equation}
Z^{\RI}_\Gamma(\mu a)=
\left(1 - \dfrac{\alpha_s^2(\mu)}{\left( 4\pi \right) ^2}\Delta _q^{(2)}\right)
Z^{'\RI}_\Gamma(\mu a)\; ,
\end{equation}
which satisfy the Ward identities at the NNLO.
The matching between RI and $\MSbar,NDR$ requires continuum perturbation 
theory only \cite{NPM,eps'/eps}.
Since both RI and $\MSbar,NDR$ respect chirality and the renormalized operators
with the correct chiral behaviour are unique, 
we have $Z_A^{\RI} = Z_A^{\msbar}$ and $Z_V^{\RI} = Z_V^{\msbar}$. 
For the same reason 
\begin{equation}
\frac{Z_S^{\RI}}{Z_S^{\msbar}} = \frac{Z_P^{\RI}}{Z_P^{\msbar}} = 
\frac{Z_m^{\msbar}}{Z_m^{\RI}}\; ,
\label{eq:WIrel} 
\end{equation}  
where $Z_m$ is the quark mass renormalization.
In ref.~\cite{mass_NNLO} $Z_m^{\msbar}/Z_m^{\RI}$ has been computed
up to order $\alpha_s^2$. Using eq.~(\ref{eq:WIrel})
one can compute at the same order the matching coefficients of the scalar 
and pseudoscalar densities 
\begin{equation}
\label{eq:matching}
Z_\Gamma^{\msbar}(\mu)= \left(1+ 
\dfrac{\alpha_s(\mu)}{4\pi}C^{(1)}
+\dfrac{\alpha_s^2(\mu)}{(4\pi) ^2}C^{(2)} \right)
Z_\Gamma^{\RI}(\mu), 
\label{eq:Z_MSbar}
\end{equation} 
where $\Gamma=P,S$ and in the Landau gauge,
\begin{eqnarray}
\label{eq:zmri}
C^{(1)} & = & 
  {{8 \left(N_c^2 - 1 \right) }\over {4\,N_c}}\, , \nonumber \\
C^{(2)} & = & 
  {{\left(N_c^2 -1 \right) }\over {96\,N_c^2}} \,
  \left( -309 + 3029\,N_c^2
\right.  \\
& \phantom{x} & \phantom{xxxxxxxx}
\left. - 288\,\zeta_3 - 576\,N_c^2\,\zeta_3 - 356\,N_c\,n_f  \right)\ . 
\nonumber
\end{eqnarray}
where $\zeta_3=1.20206\ldots$. 
The dependence on the gauge and the external states of the
RI scheme will cancel with the corresponding dependence of the matching
coefficients in eq.~(\ref{eq:Z_MSbar}), 
up to higher orders in continuum perturbative expansion and 
up to discretization errors.

\section{Renormalization Group Analysis}
\label{sec:RGE}
The RGE expresses a general property of the 
Green's functions of a renormalized theory and therefore they are valid
non-perturbatively. To study the RG properties of bilinears we work in 
the $\MSbar$ scheme and in the Landau gauge.
The generic, forward, renormalized two-point Green's 
function, computed between quark states of virtuality $p^2$ obeys the RGE
\begin{equation}
\label{eq:rge}
\left[ \mu ^2\frac d{d\mu ^2}+\frac{\gamma_\Gamma}2\right] \Gamma\lp \frac{p}
{\mu} \rp =\left[ \mu ^2\frac \partial{\partial\mu ^2} + 
\beta \left( \alpha_s\right) \frac \partial {\partial \alpha_s}
+\frac{\gamma_\Gamma} 2\right] \Gamma\lp \frac{p}{\mu} \rp =0\; , 
\end{equation}
where the QCD $\beta $-function and the anomalous dimension of the
renormalized operator $\hat O_\Gamma$ are gauge invariant to all orders in
perturbation theory and are defined as:
\begin{eqnarray}
\frac{\beta (\alpha_s)}{4\pi}&=& \mu^2\frac{d}{d\mu ^2} \lp \frac{\alpha_s}{4\pi} \rp \, 
= \, - \sum_{i=0}^\infty \beta _i\left( \frac{\alpha_s}{4\pi } \right)^{i+2},\\
\gamma_\Gamma (\alpha_s) &=& - 2Z_\Gamma^{-1}\mu ^2\frac d{d\mu ^2}Z_\Gamma=
\sum_{i=0}^\infty\gamma_\Gamma^{(i)}\left( \frac{\alpha_s}{4\pi }\right)^{i+1}.
\end{eqnarray}
In a continuum regularization which respects chirality the axial and vector 
currents do not get renormalized, i.e. $Z_A=Z_V=1$, as can
be easily shown through the Ward Identities that they satisfy. 
Since $m(\mu)P(\mu)$, with $P(\mu)$ the pseudoscalar density, is
renormalization group invariant, the scalar and pseudoscalar densities have
renormalization constants which obey $Z_S=Z_P=1/Z_m$\footnote{We stress that
we are studying the RGE evolution in a continuum renormalization scheme.}. 
Therefore one can express the anomalous dimension $\gamma_\Gamma$ of
bilinear operators as a function of $\gamma_m$:
\begin{eqnarray}
\gamma_A & = & \gamma_V = 0\; ,\\
\gamma_P & = & \gamma_S = - \gamma_m\; .\nonumber   
\label{eq:AWIg}
\end{eqnarray}
To solve the RGE's in the NNLO approximation, the expansions of the 
$\beta$ function and anomalous dimension up to three loops is required.  

The running of the coupling constant $\alpha_s^{\msbar}$ is given by
\begin{eqnarray}
\label{alphaeff}
     \frac{\alpha_s^{\msbar}}{4\pi}(q^2) & = &
        \frac{1}{\beta_0 \ln(q^2)}
        - \frac{\beta_1}{\beta_0^3}
        \frac{\ln\  \ln(q^2)}{\ln^2(q^2)}
  \nonumber \\
     && + \frac{1}{\beta_0^5 \ln^3(q^2)}\left( \beta_1^2 \ln^2\ln(q^2)
   - \beta_1^2 \ln \ \ln(q^2)+ \beta_2^{\msbar}\beta_0 - \beta_1^2 \right)\; ,
\end{eqnarray}
where $q^2=(\mu/\Lambda_{\QCD}^{\msbar})^2$.  For the continuum 
$\MSbar$ scale parameter $\Lambda_{\QCD}^{\msbar}$ at the NNLO, 
in the quenched approximation, we have
used $\Lambda_{\QCD}^{\msbar}=0.251\pm 0.021$~GeV \cite{m_q_SF}. 
The QCD $\beta$-function is scheme independent only up to two loops.
The additional term of the expansion has been computed in the $\MSbar$ scheme 
in \cite{b4}:
\begin{eqnarray}
\beta _0 & = & \frac{11}{3}N_c - \frac{2}{3} n_f, \nonumber \\
\beta _1 & = & \frac{34}3N_c^2-\frac{10}3N_cn_f-\frac{(N_c^2-1)}{N_c}n_f, \\
\beta _2^{\msbar} & = & \frac{2857}{54}N_c^3+\frac{\left(N_c^2-1\right) ^2}{
4 N_c^2}n_f-\frac{205}{36}\left( N_c^2-1\right) n_f 
\nonumber \\
&\phantom{x}&\;-\frac{1415}{54}N_c^2n_f+\frac{11}{18}\frac{\left(N_c^2-
1\right) }{N_c}n_f^2+\frac{79}{54}N_cn_f^2 \nonumber 
\label{eq:beta}
\end{eqnarray}
The mass anomalous dimension in the $\MSbar$ scheme up to three loops 
is given by \cite{g4a,g4b}:
\begin{eqnarray}
\gamma_m^{(0)} &=& 3\frac{N_c^2-1}{N_c},  \nonumber \\
\gamma_m^{(1)} &=& \frac{N_c^2-1}{N_c^2}\left( -\frac 34+%
\frac{203}{12}N_c^2-\frac 53N_cn_f\right),  \\
\gamma_m^{(2)} &=& \frac{N_c^2-1}{N_c^3}\left[ \frac{129}8-%
\frac{129}8N_c^2+\frac{11413}{108}N_c^4\right.   \nonumber \\
&\phantom{x}&\;\left. +n_f\left( \frac{23}2N_c-\frac{1177}{54}%
N_c^3-12N_c\zeta _3-12N_c^3\zeta _3\right) -\frac{35}{27}N_c^2n_f^2\right] ,
\nonumber  \label{eq:gam}
\end{eqnarray}
where $\zeta$ is the Riemann zeta function. 
 
The evolution of the renormalized bilinear operators is determined by 
eqs.~(\ref{eq:rge}) and (\ref{eq:AWIg}).
The solution can be expressed in the $\MSbar$ scheme in the form \cite{Tarrach}
\begin{equation}
\label{eq:msol}
\hat O_\Gamma^{\msbar}( \mu ) 
=\frac{c_\Gamma^{\msbar}(\mu)}{c_\Gamma^{\msbar}(\mu_0)}
 \hat O_\Gamma^{\msbar}( \mu _0 ), 
\end{equation}
where
\begin{eqnarray}
c_\Gamma^{\msbar} \left( \mu \right) &=&\alpha_s \left( \mu \right)^{\overline{\gamma }
^{(0)}_\Gamma}\left\{ 1 +\frac{\alpha_s}{4\pi } 
\left( \overline{\gamma }^{(1)}_\Gamma-
\overline{\beta }_1 \overline{\gamma }^{(0)}_\Gamma\right) \right.  \\
&+&\left. \frac 12\left( \frac{\alpha_s \left( \mu \right)}{4\pi }\right) ^2
\left[
\left( \overline{\gamma }^{(1)}_\Gamma-\overline{\beta }_1\overline{\gamma
}^{(0)}_\Gamma\right) 
^2+ \overline{\gamma }^{(2)}_\Gamma + \overline{\beta }_1^2\overline{\gamma
}^{(0)}_\Gamma-
\overline{\beta }_1\overline{\gamma }^{(1)}_\Gamma-\overline{\beta }_2
\overline{\gamma 
}^{(0)}_\Gamma \right] \right\}, \nonumber  
\label{eq:calfa}
\end{eqnarray}
with $\overline{\beta }_i=\beta _i/\beta _0$ and 
$\overline{\gamma }^{(i)}_\Gamma=\gamma_\Gamma^{(i)}/\left( 2\beta _0\right) $.\\
By using the eqs.~(\ref{eq:WIrel}) and (\ref{eq:msol}), the evolution 
of the bilinear quark operators at the NNLO in the RI scheme becomes
\begin{equation}
\hat O_\Gamma^{\RI}( \mu ) 
=\frac{c_\Gamma^{\RI}(\mu)}{c_\Gamma^{\RI}(\mu_0) }
\hat O_\Gamma^{\RI}(\mu_0), 
\label{eq:evol_RI}
\end{equation}
where
\begin{equation}
c_\Gamma^{\RI}\left( \mu \right) 
=\frac{Z_\Gamma^{\RI}(\mu)}{Z_\Gamma^{\msbar}(\mu) } 
\ c_\Gamma^{\msbar} \left( \mu \right)\; . 
\label{eq:c_RI}
\end{equation}

Eqs.~(\ref{eq:evol_RI}) and (\ref{eq:c_RI}) define the evolution at NNLO in the
RI scheme of the renormalization constants with the scale $\mu$.
In order to compare with the numerical NP results we define a 
Renormalization Group Invariant (RGI) constant as 
\begin{equation}
\label{eq:Z_RGI} 
Z_\Gamma^{\RGI}(a)=\frac{Z^{'\RI}_\Gamma(\mu a,m_qa=0)}{c_\Gamma^{'\RI}(\mu)},
\end{equation}
where
\begin{equation}
c_\Gamma^{'\RI}(\mu) = 
\frac{Z_\Gamma^{'\RI}(\mu)}{Z_\Gamma^{\RI}(\mu)} 
\ c_\Gamma^{\RI}(\mu)\; 
\end{equation}
takes into account the mismatch between $Z_q$ and $Z'_q$, 
cf.~eq~(\ref{eq:deltaq1}).  Up to higher order terms in continuum perturbation
theory and up to discretization errors, $Z_\Gamma^{\RGI}(a)$ should be 
independent of $\mu$, in the region in which  perturbation theory is valid,
i.e.~ $\mu\gtapprox 2$ GeV, independent of the renormalization scheme,
of the external states and gauge invariant.
Being the continuum evolution already at NNLO, 
we assume any scale dependence to be dominated by discretization effects.
As an estimate of this systematic error we will take the semidispersion of the 
values of the renormalization constants in the perturbative region.

\section{Non-perturbative renormalization constants}
\label{sec:Z}

The renormalization constants for the bilinears presented in this paper
are obtained at three different gauge couplings $g_0^2$, corresponding to 
$\beta=6/g_0^2=6.0,6.2$ and $6.4$ using both the standard Wilson action and the
tree-level improved SW-Clover fermion action \cite{SW}\footnote{We note that
in our implementation of the tree-level improvement program the relationship 
$Z_S=1/Z_m$ is not true anymore because we have used non-local 
``$\dslash$--rotated'' operators; we refer to \cite{NPM,m_q_APE}
for details.}. A summary of the parameters used in the 
NP calculation of the renormalization constants
is presented in tab.~\ref{tab:params_Z}.
The errors have been obtained with the jacknife method, decimating 10 
configurations at a time.  The lattice scale $a^{-1}$ for the different
couplings has been determined from $M_{K^*}$ \cite{Allton} 
and is shown in tab.~\ref{tab:params_matrix}.

\begin{table}[t]
\centering
\begin{tabular}{|c|cccccc|}
\hline\hline
$\beta$ & 6.0  & 6.0    & 6.2 & 6.2    & 6.4 & 6.4  \\
Action   & SW   & Wilson & SW  & Wilson & SW  & Wilson \\
\# Confs & 100  & 100    & 180 & 100    & 60  & 60     \\
Volume   &$16^3\times 32$ & $16^3\times 32$ & $16^3\times 32$ & $16^3\times 32$
         &$24^3\times 32$ & $24^3\times 32$ \\
\hline
$\kappa$ &0.1425& 0.1530 & 0.14144 & 0.1510 & 0.1400 & 0.1488 \\
         &0.1432& 0.1540 & 0.14184 & 0.1515 & 0.1403 & 0.1492 \\
         &0.1440& 0.1550 & 0.14224 & 0.1520 & 0.1406 & 0.1496 \\
         &      &        & 0.14264 & 0.1526 & 0.1409 & 0.1500 \\
\hline
$\kappa_{\crit}$ &0.14551& 0.15683 & 0.14319 & 0.15337 & 0.14143 & 0.15058 \\
\hline\hline
\end{tabular}
\caption{Summary of parameters used in the non-perturbative calculation 
of the renormalization constants.}
\label{tab:params_Z}
\end{table}

\subsection{Axial and vector currents}

\begin{figure}[t] 
\vspace{-4.0truecm}
\begin{center}
\setlength{\epsfxsize}{20cm}
\setlength{\epsfysize}{20cm}
\epsfig{figure=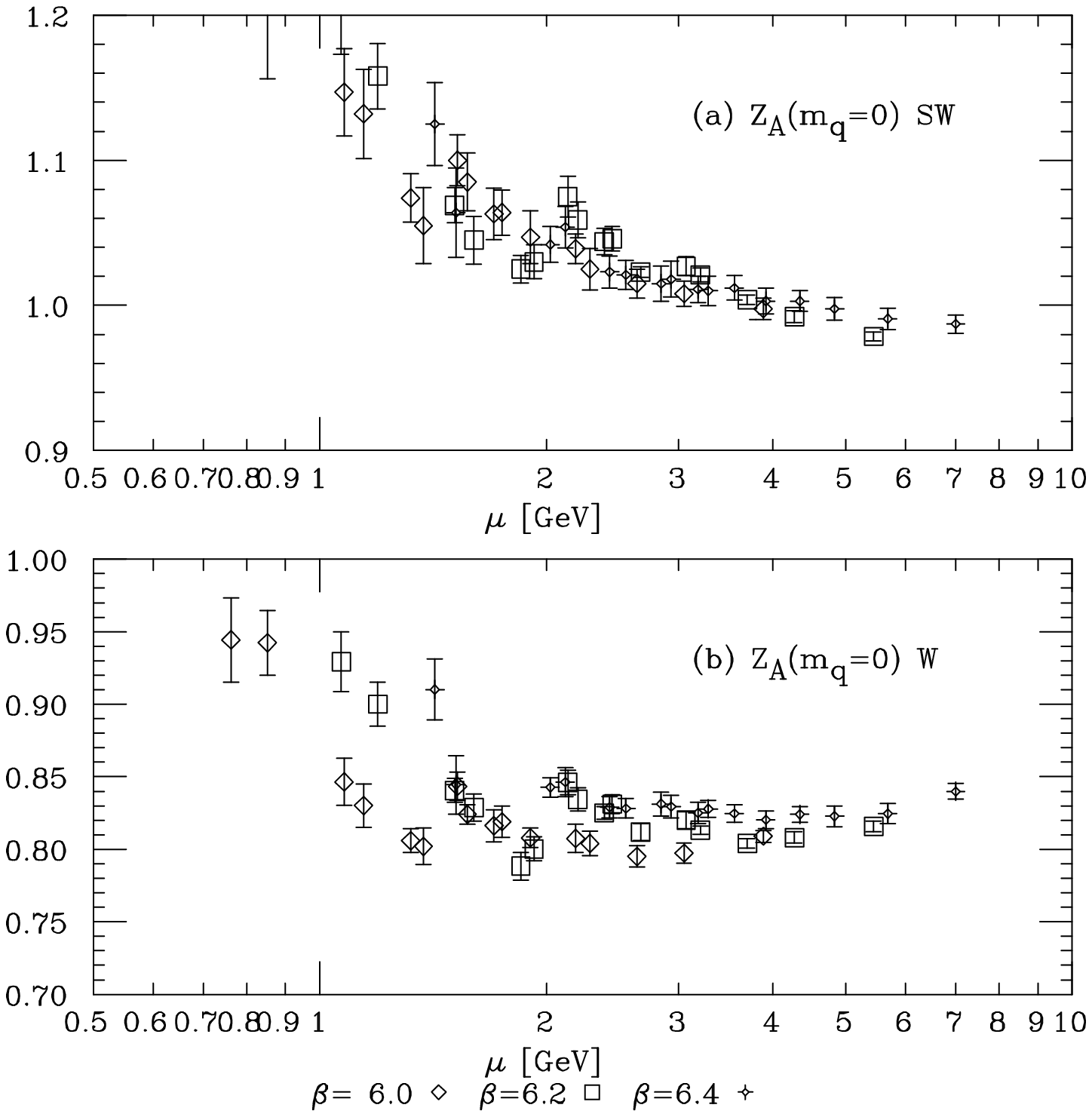,height=20cm,angle=0}
\vspace{-4.5truecm}
\end{center}
\caption{Renormalization of the axial current for (a) SW and (b) Wilson 
action as a function of $\mu$ for all couplings. The lattice spacing is
determined from $M_{K^*}$.\protect\label{fig:Z_A(mu)}} 
\end{figure} 

\begin{figure}[t] 
\vspace{-4.0truecm}
\begin{center}
\setlength{\epsfxsize}{20cm}
\setlength{\epsfysize}{20cm}
\epsfig{figure=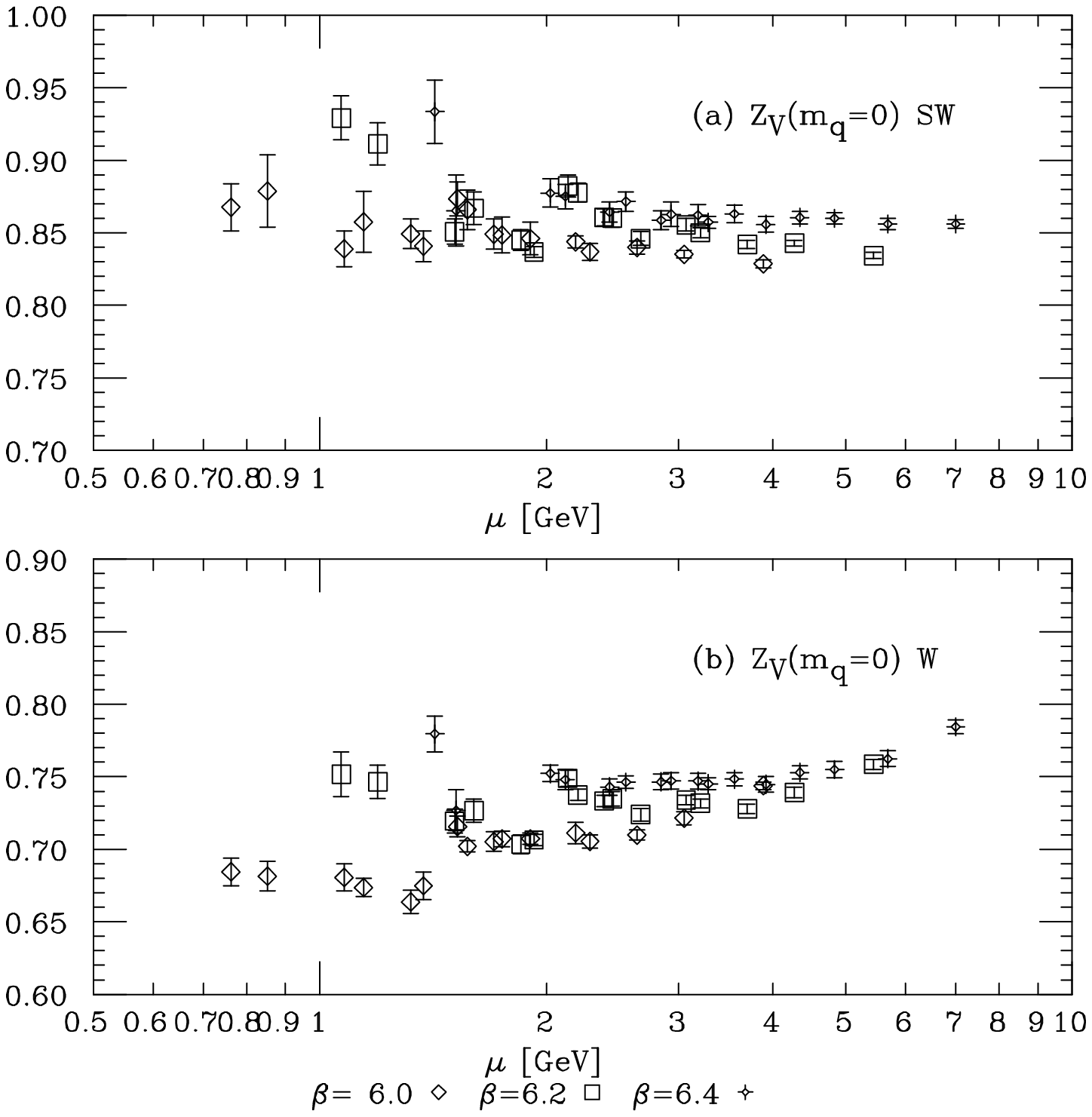,height=20cm,angle=0}
\vspace{-4.5truecm}
\end{center}
\caption{Renormalization of the vector current for (a) SW and (b) Wilson 
action as a function of $\mu$ for all couplings. The lattice spacing is
determined from $M_{K^*}$.\protect\label{fig:Z_V(mu)}}
\end{figure} 

Let us consider first the axial and vector currents.  Since each obeys a 
chiral Ward Identity \cite{Bochicchio} their renormalization constants
are finite, i.e.\ scale independent.  In figs.~\ref{fig:Z_A(mu)} and
\ref{fig:Z_V(mu)} we show $Z_A$ and $Z_V$, calculated in the RI scheme
in the chiral limit for both the SW and Wilson actions, 
as a function of the renormalization scale $\mu$.  
As explained in \cite{NPM}, we expect to find a ``window'' in the
range of values of $\mu$ in which the $Z$'s are scale independent.  
Since at large values of $\mu$ the $Z$'s will
be more sensible to discretization errors,  we expect the window to be wider 
as we approach the continuum.

We can clearly see that these expectations are satisfied by the data
shown in figs.~\ref{fig:Z_A(mu)} and \ref{fig:Z_V(mu)}, and also reported in 
tabs.~\ref{tab:Z(m=0)SW} and \ref{tab:Z(m=0)W}.  For the axial and vector 
currents, the RGI values
given in the tables are simply the values of the constants calculated
at $\mu a\simeq 1$, since $c_\Gamma^{'\RI}(\mu)$ for $Z_A$ and $Z_V$ is totally
negligible in comparison with the final error. The first error is statistical, while the second
is the systematic error estimated as the semidispersion of the
values in the region in which we believe perturbation theory to be reliable,
i.e. $\mu\gtapprox 2$ GeV.  In the SW case, the axial current shows a more
pronounced systematic effect compared to the vector current, 
reflected in a reduced stability in $\mu$ of the plots.
In the Wilson case, on the other hand,  it's the vector current that
is more fluctuating in the scale $\mu$.
Moreover, as far the comparison between the SW and Wilson actions is concerned
(at the same value of $\beta$), we note that for the axial current the Wilson
action seems more stable than the SW one, while for the vector current
it's the SW action which shows a more pronounced plateau.

In tab.~\ref{tab:Z(m=0)} we also give the values of
$Z_A$ and $Z_V$ determined from the WI's \cite{MM,Crisafulli,Henty}.  
Caution should be used in comparing the
values obtained from the NPM and the WI's as the latter are sometimes 
available only for finite values of the quark mass,  
but since the mass dependence is expected to be weak, it is still a
significant check.
The agreement between the NPM and the WI's
is good for the axial current in the SW case, around $\mu a\approx 1$,
as already found in \cite{NPM}.  On the other hand, in the Wilson case
the WI value at $\beta=6.0$ is even larger than the NPM at $\beta=6.4$,
although the error quoted from the WI is so large ($\approx 10\%$) 
that a significant comparison is not possible.
For the vector current, we notice that in SW case the NP values are
slightly higher than the WI although the discrepancy tends to diminish
as $\beta$ increases, whereas in the Wilson case, the comparison is 
more complicated.  This is due to the well-know discretization errors
which affect the hadronic matrix elements of the vector current, 
in particular of the conserved one \cite{Crisafulli}.
We expect the values of $Z_A$ and $Z_V$ to approach unity 
as move towards the continuum limit and the coupling goes to zero.  
All the data obtained with the NPM support this conclusion.

In tab.~\ref{tab:Z(m=0)} we also present the 
one-loop perturbative values \cite{Zhang,Borrelli}, evaluated at 
$\mu a=1$ in the RI scheme according to eq.~(\ref{eq:Z_MSbar}).
We present both Stardard PT (SPT), in which the bare coupling $g_0^2$ is
used as the expansion parameter, and Boosted PT (BPT) in which the 
coupling used is \cite{Lepage}
\begin{equation}
\alpha_s^{\BPT}=\frac{1}{\<\Box\>}\alpha_s^{\latt}
               =\frac{1}{\<\Box\>}\frac{g_0^2}{4\pi}
\end{equation}
where $\<\Box\>$ is the expectation value of the plaquette.
In all the perturbative calculations, the effects coming from
the discretization of loop integrals on a finite lattice have
been ignored.  As already emphasized in \cite{NPM}, perturbation theory 
does not agree with the non-perturbative values, except in a number
of limited cases, as the vector current.  In the case of the pseudoscalar
density, the failure is quite dramatic and has important phenomenological
consequences on the determination of the quark masses and the quark 
condensate, as explained in refs.~\cite{m_q_APE}.
Moreover, compared to the NPM, the perturbative values show a less pronounced
variation with the coupling.   Thus, even if the agreement improves as
we approach the continuum, the discrepancy remains sizable even at the
smallest coupling.

The exploratory numerical calculations in \cite{NPM} were carried out at
$\beta=6.0$ and a single value of the hopping parameter $\kappa=0.1425$,
corresponding to $m_qa\simeq 0.07$.  We have checked that the results
are compatible with ours within statistical errors.  We 
have chosen not to separate the time and space components
of the currents as they agree within errors. 
It is worth noting that the sizable difference between the renormalization
constants of the different components and the large fluctuations found in 
\cite{NPM} at large values of
$\mu^2a^2$ were not due to strong discretization effects, but to the
choice of the external momenta.  In fact, the components of the momenta
$p=(p_0,p_1,p_2,p_3)$ were chosen with large differences among the $p_i$'s,
thus greatly enhancing the rotational symmetry breaking on the hypercubic
lattice.

Recently, the non-perturbative method of ref.~\cite{NPM} 
has also been applied to the renormalization of bilinears
in ref.~\cite{Oelrich}, in which ``momentum'' sources and sinks and
translational invariance are used to reduce the statistical noise in the
determination of the $Z$'s.  The method has been tested for Wilson fermions
at $\beta=6.0$, and the values of the $Z$'s in the chiral limit calculated 
at $\mu a\simeq 1$ are $Z_A=0.7807(8)$, $Z_V=0.6822(7)$, $Z_P=0.4357(17)$
and $Z_S=0.6808(15)$.  The errors quoted are purely statistical and
are much smaller than the systematic errors coming from the discretization
effects, which are expected to be quite large in the Wilson case.
With this in mind, we find that the agreement between the values of 
\cite{Oelrich} and ours is resonable.

\begin{table}
\centering
\begin{tabular}{|c|c|ccccc|}
\hline\hline
$\beta$ &
$\mu^2 a^2$ &$Z_A$ & $Z_V$ & $Z_P$ & $Z_S$ & $Z_P/Z_S$
\\
\hline
   & 0.308 & 1.147(30)& 0.839(12)& 0.199( 8)& 0.711(36)& 0.293(20)\\
   & 0.617 & 1.100(18)& 0.874(12)& 0.335(11)& 0.799(24)& 0.422(16)\\
   & 0.964 & 1.047(18)& 0.846(11)& 0.409( 8)& 0.834(18)& 0.492(13)\\
   & 1.272 & 1.039(10)& 0.844( 4)& 0.457( 6)& 0.862(18)& 0.531(12)\\
6.0
   & 1.388 & 1.025(14)& 0.837( 6)& 0.467( 6)& 0.871(18)& 0.537(14)\\
   & 1.851 & 1.015(10)& 0.840( 5)& 0.516( 6)& 0.905(11)& 0.570( 9)\\
   & 2.467 & 1.008( 9)& 0.835( 3)& 0.555( 4)& 0.943(12)& 0.588(10)\\
   & 4.010 & 0.998( 7)& 0.829( 3)& 0.610( 3)& 0.993(11)& 0.614( 8)\\
   &  RGI  & 1.047(18)(25)& 0.846(11)( 9)& 0.278( 5)(40)& 0.567(12)( 8)& 0.492(13)(61)\\
\hline
   & 0.308 & 1.069(12)& 0.851( 9)& 0.248( 8)& 0.713(28)& 0.372(20)\\
   & 0.617 & 1.075(14)& 0.882( 8)& 0.401( 9)& 0.837(21)& 0.496(17)\\
   & 0.964 & 1.023( 4)& 0.846( 5)& 0.466( 4)& 0.851(11)& 0.563( 8)\\
   & 1.272 & 1.027( 7)& 0.856( 4)& 0.523( 4)& 0.904( 7)& 0.581( 6)\\
6.2
   & 1.388 & 1.021( 5)& 0.850( 4)& 0.527( 4)& 0.906(10)& 0.590( 7)\\
   & 1.851 & 1.004( 3)& 0.842( 3)& 0.564( 3)& 0.928( 3)& 0.608( 4)\\
   & 2.467 & 0.992( 4)& 0.843( 2)& 0.602( 3)& 0.950( 6)& 0.630( 6)\\
   & 4.010 & 0.978( 3)& 0.834( 2)& 0.642( 2)& 0.983( 4)& 0.654( 5)\\   
   &  RGI  & 1.023( 4)(24)& 0.846( 5)(11)& 0.295( 2)(32)& 0.540( 7)( 9)& 0.563( 8)(45)\\
\hline
   & 0.313 & 1.042(12)& 0.877(10)& 0.405(11)& 0.739(15)& 0.549(18)\\
   & 0.617 & 1.015(12)& 0.859( 6)& 0.498( 8)& 0.810(17)& 0.616(16)\\
   & 0.964 & 1.012( 9)& 0.863( 6)& 0.555( 6)& 0.852(13)& 0.652(10)\\
   & 1.169 & 1.003( 9)& 0.856( 5)& 0.572( 5)& 0.869(15)& 0.659(11)\\
6.4
   & 1.439 & 1.003( 7)& 0.861( 4)& 0.597( 6)& 0.896(15)& 0.667(12)\\
   & 1.782 & 0.997( 8)& 0.860( 4)& 0.618( 5)& 0.918(13)& 0.673(11)\\
   & 2.467 & 0.991( 7)& 0.856( 4)& 0.646( 4)& 0.946(13)& 0.683( 9)\\
   & 3.740 & 0.987( 6)& 0.856( 3)& 0.676( 4)& 0.974(10)& 0.694( 9)\\
   &  RGI  & 1.012( 9)(12)& 0.863( 6)( 4)& 0.327( 4)(17)& 0.502( 8)( 9)& 0.652(10)(21)\\
\hline\hline
\end{tabular}
\caption{Non-perturbative values of $Z_\Gamma^{\RI}(m_q=0)$ with the SW 
action, for all couplings at several renormalization scales 
$\mu^2a^2$. For the values at different scales the errors are  statistical.
The RGI values are computed from that one at $\mu a \simeq 1$ according to 
the eq.~(\ref{eq:Z_RGI}) and the first error is statistical, 
the second systematic as explained in text.}
\label{tab:Z(m=0)SW}
\end{table}

\begin{table}
\centering
\begin{tabular}{|c|c|ccccc|}
\hline\hline
$\beta$ &
$\mu^2 a^2$ &$Z_A$ & $Z_V$ & $Z_P$ & $Z_S$ & $Z_P/Z_S$
\\
\hline
  & 0.308 & 0.847(16)& 0.681( 9)& 0.241( 6)& 0.559(13)& 0.445(10)\\
  & 0.617 & 0.843(10)& 0.716( 7)& 0.372( 3)& 0.654( 7)& 0.575(10)\\
  & 0.964 & 0.808( 7)& 0.707( 4)& 0.447( 5)& 0.682( 9)& 0.657( 8)\\
  & 1.272 & 0.807(10)& 0.711( 7)& 0.492( 6)& 0.713(10)& 0.692(10)\\
6.0
  & 1.388 & 0.804( 8)& 0.705( 5)& 0.503( 4)& 0.725(10)& 0.695(10)\\
  & 1.851 & 0.795( 7)& 0.710( 4)& 0.549( 3)& 0.746(10)& 0.736( 9)\\
  & 2.467 & 0.797( 7)& 0.722( 5)& 0.588( 3)& 0.762( 7)& 0.772( 7)\\
  & 4.010 & 0.809( 4)& 0.744( 3)& 0.649( 2)& 0.795( 5)& 0.816( 4)\\
  &  RGI  & 0.808( 7)( 7)& 0.707( 4)(19)& 0.294( 3)(39)& 0.448( 6)( 5)& 0.657( 8)(80)\\
\hline
  & 0.308 & 0.841( 8)& 0.719( 8)& 0.272( 8)& 0.615(21)& 0.463(25)\\
  & 0.617 & 0.846( 8)& 0.749( 6)& 0.436(11)& 0.708( 8)& 0.629(19)\\
  & 0.964 & 0.812( 6)& 0.724( 4)& 0.499( 5)& 0.722( 7)& 0.695(12)\\
  & 1.272 & 0.820( 6)& 0.734( 3)& 0.539( 4)& 0.755( 7)& 0.718( 8)\\
6.2
  & 1.388 & 0.813( 3)& 0.732( 3)& 0.548( 5)& 0.755( 8)& 0.742(10)\\
  & 1.851 & 0.804( 3)& 0.728( 3)& 0.580( 3)& 0.768( 5)& 0.761( 6)\\
  & 2.467 & 0.808( 4)& 0.739( 4)& 0.621( 3)& 0.787( 4)& 0.790( 5)\\
  & 4.010 & 0.816( 4)& 0.759( 3)& 0.675( 3)& 0.810( 4)& 0.833( 4)\\
  &  RGI  & 0.812( 6)( 8)& 0.724( 4)(17)& 0.310( 3)(31)& 0.448( 5)( 5)& 0.695(12)(69)\\
\hline
   & 0.313 & 0.843( 7)& 0.752( 6)& 0.413( 7)& 0.649( 9)& 0.643(14)\\
   & 0.617 & 0.831( 8)& 0.747( 6)& 0.516( 5)& 0.710(12)& 0.730(12)\\
   & 0.964 & 0.825( 6)& 0.748( 5)& 0.572( 4)& 0.742( 7)& 0.773( 8)\\
   & 1.169 & 0.820( 6)& 0.745( 6)& 0.590( 5)& 0.754( 6)& 0.785( 7)\\
6.4
   & 1.439 & 0.824( 6)& 0.753( 4)& 0.617( 5)& 0.774( 6)& 0.797( 7)\\
   & 1.782 & 0.823( 7)& 0.755( 6)& 0.637( 5)& 0.787( 7)& 0.810( 6)\\
   & 2.467 & 0.825( 7)& 0.762( 5)& 0.669( 5)& 0.803( 7)& 0.833( 4)\\
   & 3.740 & 0.840( 5)& 0.784( 5)& 0.710( 5)& 0.830( 6)& 0.855( 3)\\
   &  RGI  & 0.825( 6)(10)& 0.748( 5)(20)& 0.333( 2)(21)& 0.431( 4)( 4)& 0.773( 8)(41)\\
\hline\hline
\end{tabular}
\caption{Non-perturbative values of $Z_\Gamma^{\RI}(m_q=0)$ with the Wilson action, 
for all couplings at several renormalization scales $\mu^2a^2$.
For the values at different scales the errors are  statistical.
The RGI values are computed from that one at $\mu a \simeq 1$ according to 
the eq.~(\ref{eq:Z_RGI}) and the first error is statistical, 
the second systematic as explained in text.}
\label{tab:Z(m=0)W}
\end{table}

\begin{table}
\centering
\begin{tabular}{|c||c|c|ccccc|}
\hline\hline
Action &
$\beta$ &
Method &$Z_A$ & $Z_V$ & $Z_P$ & $Z_S$ & $Z_P/Z_S$
\\
\hline
& 6.0  & SPT                  & 0.98    & 0.90    & 0.74     & 0.91    & 0.82 \\ 
& 6.0  & BPT                  & 0.97    & 0.83    & 0.56     & 0.84    & 0.67 \\ 
& 6.0  & WI \cite{Crisafulli} & 1.10(2) & 0.80(2) &          &         & 0.61(2)\\ 
& 6.0  & [This work]          & 1.05(3) & 0.85(1) & 0.41(6)  & 0.83(2) & 0.49(6)\\ 
\hline
SW
& 6.2  & SPT                  & 0.98    & 0.90    & 0.75    & 0.91    & 0.83\\ 
& 6.2  & BPT                  & 0.97    & 0.84    & 0.59    & 0.85    & 0.70\\ 
& 6.2  & WI \cite{Henty}      & 1.05(1) & 0.82(1) &         &         & 0.69(4) \\ 
& 6.2  & [This work]          & 1.02(2) & 0.85(1) & 0.47(5) & 0.85(2) & 0.56(5) \\ 
\hline
& 6.4  & SPT                  & 0.98    & 0.91    & 0.76    & 0.91    & 0.83   \\ 
& 6.4  & BPT                  & 0.97    & 0.85    & 0.62    & 0.86    & 0.72   \\ 
& 6.4  & [This work]          & 1.01(1) & 0.863(7)& 0.55(3) & 0.85(2) & 0.65(2)\\ 
\hline\hline
& 6.0  & SPT                  & 0.87    & 0.83    & 0.78    & 0.86    & 0.91    \\ 
& 6.0  & BPT                  & 0.78    & 0.71    & 0.62    & 0.76    & 0.82    \\ 
& 6.0  & WI \cite{MM}         & 0.85(7) & 0.79(4) &         &         &         \\ 
& 6.0  & [This work]          & 0.81(1) & 0.71(2) & 0.45(6) & 0.68(1) & 0.66(8) \\ 
\hline
Wilson
& 6.2  & SPT                  & 0.87    & 0.83    & 0.78    & 0.86    & 0.91    \\ 
& 6.2  & BPT                  & 0.79    & 0.73    & 0.65    & 0.73    & 0.84    \\ 
& 6.2  & [This work]          & 0.81(1) & 0.72(2) & 0.50(5) & 0.72(1) & 0.69(7) \\ 
\hline
& 6.4  & SPT                  & 0.87    & 0.84    & 0.79    & 0.86    & 0.91   \\ 
& 6.4  & BPT                  & 0.80    & 0.74    & 0.67    & 0.79    & 0.85   \\ 
& 6.4  & WI \cite{Crisafulli} &         & 0.71(1) &         &         &        \\ 
& 6.4  & [This work]          & 0.82(1) & 0.75(2) & 0.57(4) & 0.74(1) & 0.77(5) \\ 
\hline\hline
\end{tabular}
\caption{Perturbative values (Standard and Boosted PT) 
of $Z_\Gamma^{\RI}(m_q=0)$ and non-perturbative
values from the WI with the SW and Wilson action.  The perturbative $Z$'s are
evaluated at $\mu^2a^2=1$.}
\label{tab:Z(m=0)}
\end{table}

\subsection{Pseudoscalar and scalar densities}

\begin{figure}[t] 
\vspace{-4.0truecm}
\begin{center}
\setlength{\epsfxsize}{20cm}
\setlength{\epsfysize}{20cm}
\epsfig{figure=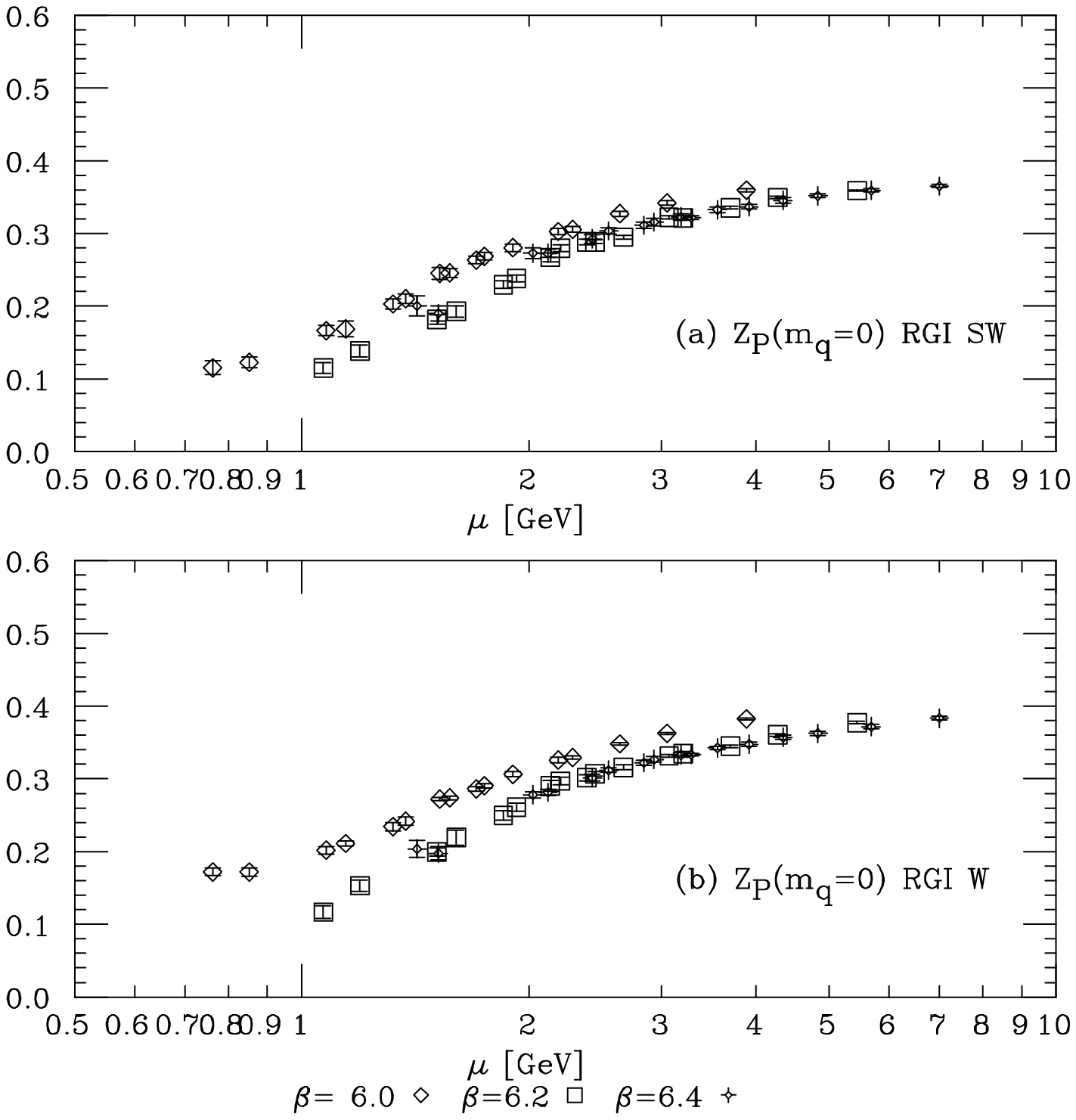,height=20cm,angle=0}
\vspace{-4.5truecm}
\end{center}
\caption{RGI values for the renormalization of the pseudoscalar density 
for (a) SW and (b) Wilson action as a function of $\mu$ for all couplings.
The lattice spacing is
determined from $M_{K^*}$.\protect\label{fig:Z_P_RGI(mu)}}
\end{figure} 

\begin{figure}[t] 
\vspace{-4.0truecm}
\begin{center}
\setlength{\epsfxsize}{20cm}
\setlength{\epsfysize}{20cm}
\epsfig{figure=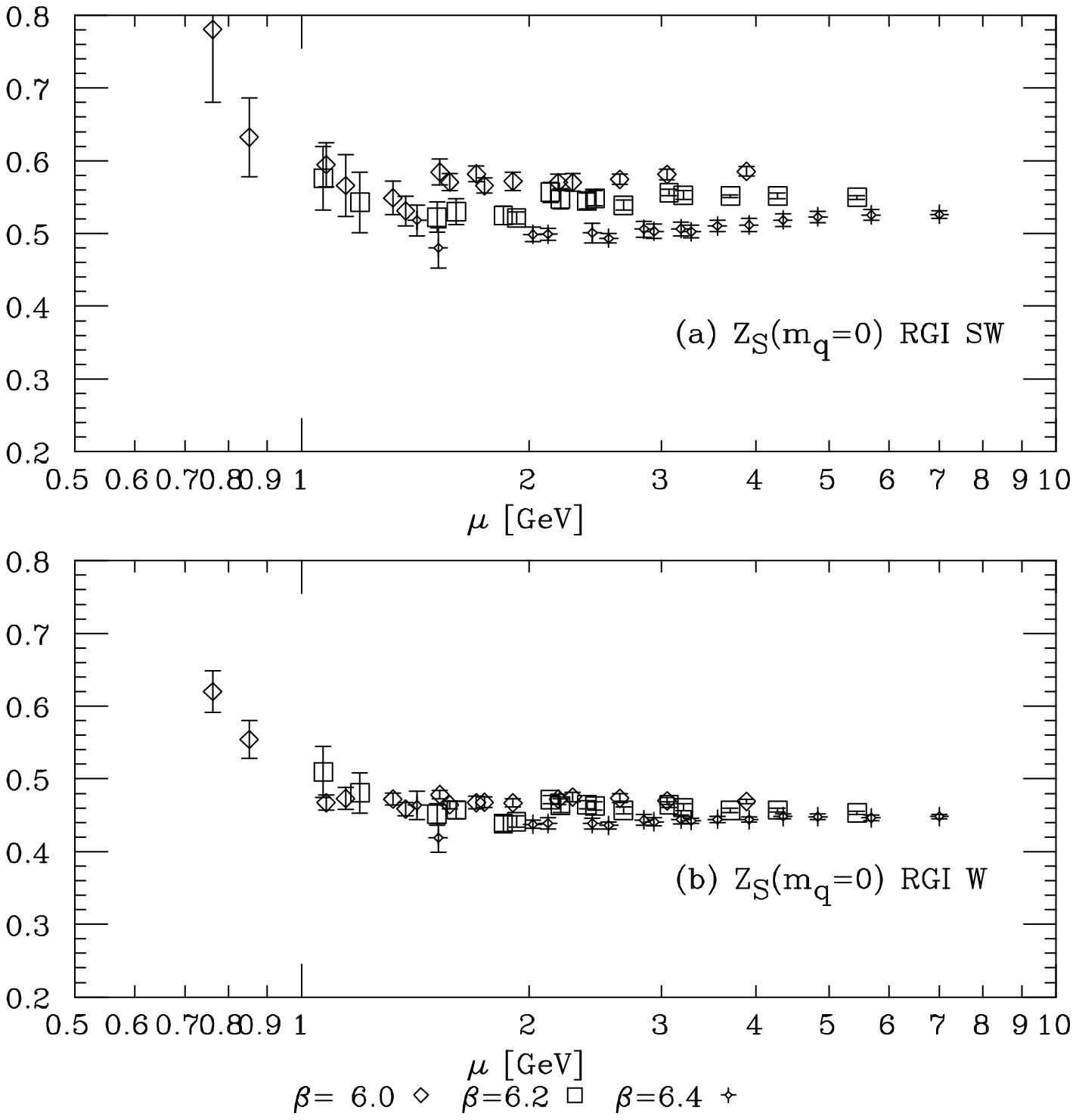,height=20cm,angle=0}
\vspace{-4.5truecm}
\end{center}
\caption{RGI values for the renormalization of the scalar density 
for (a) SW and (b) Wilson action as a function of $\mu$ for all couplings.
The lattice spacing is
determined from $M_{K^*}$.
\protect\label{fig:Z_S_RGI(mu)}}
\end{figure} 

\begin{figure}[t] 
\vspace{-4.0truecm}
\begin{center}
\setlength{\epsfxsize}{20cm}
\setlength{\epsfysize}{20cm}
\epsfig{figure=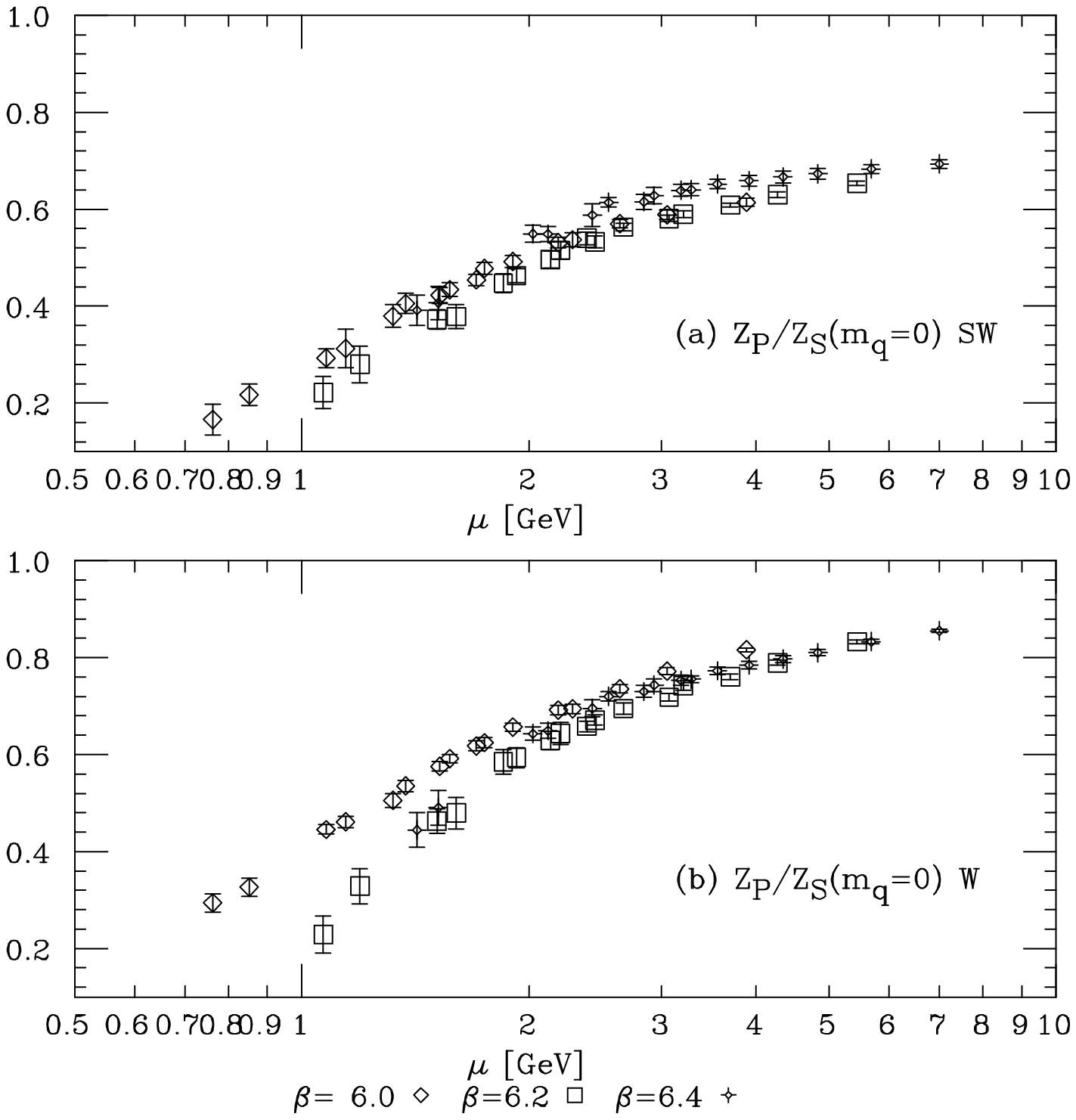,height=20cm,angle=0}
\vspace{-4.5truecm}
\caption{Renormalization of the pseudoscalar to scalar density ratio 
for (a) SW and (b) Wilson 
action as a function of $\mu$ for all couplings.
The lattice spacing is determined from $M_{K^*}$. 
\protect\label{fig:Z_PvsZ_S_RGI}  }
\end{center}
\end{figure} 

The pseudoscalar and scalar densities differ from the axial and vector
currents in that their renormalization is not finite but
scale dependent.  We show in figs.~\ref{fig:Z_P_RGI(mu)} and 
\ref{fig:Z_S_RGI(mu)} the RGI values, computed according to 
eq.~(\ref{eq:Z_RGI}).  We stress that the RGI values are only reliable
from values of $\mu\gtapprox 2$ GeV, i.e. when continuum perturbation
theory is to be trusted.
The RGI value in tabs.~\ref{tab:Z(m=0)SW} and \ref{tab:Z(m=0)W} 
has been calculated from $\mu a\simeq 1$. 
The systematic error is calculated in the same fashion as for the axial 
and vector currents.
It is easily noticeable the difference in the behaviour of $Z_P$ and $Z_S$
as a function of $\mu$.  The former approaches a plateau only at
relatively large values of $\mu$, while the latter presents a very
clear and long plateau.  This behaviour, also reflected in a much
larger systematic error for the pseudoscalar, does not depend on the choice
of the action.  

While $Z_P$ and $Z_S$ cannot be determined separately by imposing a
Ward Identity, their ratio $Z_P/Z_S$ can as it is scale independent.
For this reason it can be treated in a similar fashion to $Z_A$ and $Z_V$
and can be compared with the values obtained from the WI.
In fig.~\ref{fig:Z_PvsZ_S_RGI} we show $Z_P/Z_S$ as function of $\mu$.  
As expected, the behaviour of the ratio
is dominated by the behaviour of $Z_P$ and agreement with the 
WI determination seems to be obtained
at rather high values of the renormalization scale, at which discretization
effects should dominate.



\section{Meson decay constants}
\label{sec:matrix}

There are several interesting phenomenological quantities which can be 
extracted from the matrix elements of quark bilinears, such as 
leptonic decay constants and quark masses.
The problem of a non-perturbative measurement of quark masses has been 
addressed in \cite{m_q_APE} to which we refer the reader for all details.
A very interesting by-product of both quark masses and decay constants
is the estimate of the chiral condensate, which is of great
phenomenological relevance.  The issue will be addressed in a forthcoming
paper \cite{chi}, together with a detailed theoretical analysis.

In the remainder of this section, we concentrate on the determination
of the leptonic decay constants of mesons.
The pseudoscalar and vector decay constants, $f_{PS}$ and $f_V$, are defined as
\begin{eqnarray}
\< 0|A_0|PS \> & = & i \frac{f_{PS}}{Z_A} M_{PS}\; ,\\
\< 0|V_i|V,r \> & = & \epsilon^r_i \frac{M_V^2}{f_V Z_V}\; ,
\end{eqnarray}
where $\epsilon^r_i$ is the vector-meson polarization, $M_{PS}$ and $M_V$ are
the pseudoscalar and vector masses, $A_0$ and $V_i$ the temporal and spatial
components of the axial and vector currents respectively, and $Z_{V,A}$ the 
corresponding renormalization constants.

In tab.~\ref{tab:params_matrix} we summarize the parameters used in the
simulations.  The bare lattice decay constants have been extracted
from the appropriate correlation functions as described in 
ref.~\cite{Allton}, to which we refer for details.
In tab.~\ref{tab:results_bare} and \ref{tab:results_renorm} we present the 
results for the decay constants, both the bare unrenormalized values in 
lattice units and the renormalized ones in physical units, obtained
using the renormalization constants at $\mu a\simeq 1$.  In the estimate
of the error on renormalized decay constants we have neglected
the errors on the renormalization constants.
The non-perturbatively renormalized values
and the experimental values present roughly a $10-15\%$ 
discrepancy, which is to be expected, considering that we have performed
all calculations in the quenched approximation.  It is important, though,
to note that
compared to the values obtained with perturbative renormalization in 
ref.~\cite{Allton}, the values obtained with a 
non-perturbative renormalization are in general closer to the
experimental results.
We have not attempted an extrapolation to the
continuum limit as the physical volume at the smallest coupling is
too small to confidently extract hadronic matrix elements.  

\begin{table}[hbt]
\begin{tabular}{|c|cccccc|}
\hline\hline        
$\beta$ &6.0 &6.0&6.2 &6.2 &6.4&6.4\\ 
Action  &SW  &Wilson&SW  &Wilson &SW &Wilson\\
\# Confs&490 &320&250 &250 &400&400\\
Volume  &$18^3\times 64$&$18^3\times 64$&$24^3\times 64$&
         $24^3\times 64$&$24^3\times 64$&$24^3\times 64$\\
\hline
$\kappa$ &0.1425& 0.1530 & 0.14144 & 0.1510 & 0.1400 & 0.1488 \\
         &0.1432& 0.1540 & 0.14184 & 0.1515 & 0.1403 & 0.1492 \\
         &0.1440& 0.1550 & 0.14224 & 0.1520 & 0.1406 & 0.1496 \\
         &      &        & 0.14264 & 0.1526 & 0.1409 & 0.1500 \\

\hline
$t_1 - t_2$ & 15-28 & 15-28 & 18-28 & 18-28 & 24-30 & 24-30\\   
\hline
$a^{-1}(K^*)$        
           &2.123(62)&2.258(50)&2.719(141)&2.993(94)&4.004(195)&4.149(161)\\
\hline\hline
\end{tabular}
\caption{Summary of the parameters used in the calculation of the matrix
elements.} 
\label{tab:params_matrix}
\end{table}  

\begin{table}[hbt]
\setlength{\tabcolsep}{.16pc}
\begin{tabular}{|l|cccccc|}
\hline\hline
$\beta$ &6.0 &6.0&6.2 &6.2 &6.4&6.4\\ 
Action  &SW  &Wilson&SW  &Wilson &SW &Wilson\\
\hline
$(f_K\; a)/Z_A$
     &0.0735(18)&0.0944(26)&0.0540(23)&0.0640(21)&0.0406(14)&0.0480(16)\\
$(f_\pi\; a)/Z_A$   
     &0.0661(21)&0.0878(31)&0.0470(28)&0.0568(24)&0.0370(16)&0.0438(19)\\
\hline
$1/(f_\phi Z_V)$
     & 0.348(9) & 0.451(11)& 0.332(8) & 0.417(8) & 0.284(9) & 0.363(8) \\
$1/(f_{K^*} Z_V)$
     &0.366(14) & 0.482(16)& 0.359(20)&0.446(15) &0.290(13)&0.377(12)  \\
$1/(f_{\rho} Z_V)$
     &0.384(20) &0.513(21) &0.386(33) &0.475(22) &0.297(17)&0.391(17)  \\
\hline\hline
\end{tabular}
\caption{Lattice bare decay constants for all couplings and both actions.}
\label{tab:results_bare}
\end{table}  

\begin{table}[hbt]
\setlength{\tabcolsep}{.2pc}
\begin{tabular}{|lc|cccccc|}
\hline\hline
$\beta$&Exp&6.0 &6.0&6.2 &6.2 &6.4&6.4\\ 
Action & &SW  &Wilson&SW  &Wilson &SW &Wilson\\
\hline
$f_K$         
     &0.1598&0.163(12)&0.172(6) &0.150(8) &0.155(5) &0.164(9) &0.164(8) \\
$f_\pi$       
     &0.1307&0.147(12)&0.160(7) &0.131(10)&0.138(6) &0.150(10)&0.150(10)\\
\hline
$1/f_\phi$    
     &0.23  &0.294(8) &0.319(8) &0.281(7) &0.302(6) &0.245(8) &0.271(6) \\
$1/f_{K^*}$   
        & &0.309(12)&0.341(11)&0.304(17)&0.323(11)&0.251(11)&0.282(9) \\
$1/f_{\rho}$  
     &0.28  &0.324(17)&0.363(15)&0.327(28)&0.344(16)&0.256(14)&0.293(12)\\
\hline\hline
\end{tabular}
\caption{Non-perturbatively renormalized decay constants (in GeV) for
all couplings and both actions.}
\label{tab:results_renorm} 
\end{table}  

\section{Conclusions}
\label{sec:conclusions}

In this paper we have performed a systematic study of the
renormalization of quark bilinears, in a non-perturbative fashion.
We have also analyzed the discretization effects, by performing our
calculations with two different actions, the standard Wilson action
and the tree-level improved SW-Clover action, at three different
values of the couplings.  
We have performed a RG analysis at the NNLO and defined RGI values
for the scale dependent renormalization constants.  We have also 
used this approach to estimate the systematic error induced by discretization.
Finally, we have applied our results to the calculation of the pseudoscalar
and vector decay constants and we find that the non-perturbatively
renormalized values, albeit with still sizable statistical errors,  
show a trend towards the experimental values with respect to the
ones obtained with perturbative renormalization.
With our data an extrapolation to the continuum limit is not reliable
as the physical volume at the smallest coupling is too small to confidently 
extract hadronic matrix elements.

\begin{ack} 
We warmly thank V.~Lubicz, G.~Martinelli, M.~Testa and A.~Vladikas 
for enlightening discussions.
MT acknowledges the support of PPARC through grant GR/L22744.
\end{ack}

\end{document}